\documentclass{elsart}

\usepackage{amsmath,amssymb,bm,graphicx}

\begin{document}

\begin{frontmatter}



\title{
Study of nuclear correlation effects 
via ${}^{12}{\rm C}(\vec{p},\vec{n}){}^{12}{\rm N}({\rm g.s.},1^+)$ at 296 MeV}


\author[kyushu]{T.~Wakasa}
\ead{wakasa@phys.kyushu-u.ac.jp}
\author[kyushu]{M.~Dozono}
\author[kyushu]{E.~Ihara}
\author[kyushu]{S.~Asaji}
\author[rcnp]{K.~Fujita}
\author[rcnp]{K.~Hatanaka}
\author[hosei]{M.~Ichimura}
\author[cyric]{T.~Ishida}
\author[rcnp]{T.~Kaneda}
\author[rcnp]{H.~Matsubara}
\author[kyushu]{Y.~Nagasue}
\author[kyushu]{T.~Noro}
\author[cyric]{Y.~Sakemi}
\author[cns]{Y.~Shimizu}
\author[kyushu]{H.~Takeda}
\author[rcnp]{Y.~Tameshige}
\author[rcnp]{A.~Tamii}
\author[kyushu]{Y.~Yamada}

\address[kyushu]{Department of Physics, Kyushu University,
Fukuoka 812-8581, Japan}
\address[rcnp]{Research Center for Nuclear Physics, Osaka University,
Osaka 567-0047, Japan}
\address[hosei]{Faculty of Computer and Information Sciences, 
Hosei University, Tokyo 184-8584, Japan}
\address[cyric]{Cyclotron and Radioisotope Center, Tohoku University,
Miyagi 980-8578, Japan}
\address[cns]{Center for Nuclear Study, The University of Tokyo,
Tokyo 133-0033, Japan}

\begin{abstract}
% Text of abstract
 We report measurements of the cross section and a complete
set of polarization observables for the Gamow--Teller
${}^{12}{\rm C}(\vec{p},\vec{n}){}^{12}{\rm N}({\rm g.s.},1^+)$ reaction 
at a bombarding energy of 296 MeV.
 The data are compared with distorted wave impulse approximation
calculations employing transition form factors normalized 
to reproduce the observed beta-decay $ft$ value.
 The cross section is significantly under-predicted by the calculations 
at momentum transfers $q \gtrsim $ 0.5 ${\rm fm^{-1}}$.
 The discrepancy is partly resolved by considering the non-locality 
of the nuclear mean field. However, the calculations still under-predict
the cross section 
at large momentum transfers of $q$ $\simeq$ 1.6 ${\rm fm^{-1}}$.
 We also performed calculations employing random phase 
approximation response functions and found that the 
observed enhancement can be attributed in part 
to pionic correlations in nuclei.
\end{abstract}

\end{frontmatter}

% main text
 The prediction of pion condensation
\cite{Migdal1}
has prompted extensive experimental and theoretical studies 
of nuclear spin--isospin correlations.
 Pion condensation is expected to occur in 
cool neutron stars (NS) such as 3C58 \cite{Chandra}
because pion condensation can accelerate 
the cooling of NS \cite{tsuruta}.
 It is believed that pion condensation does not 
occur in normal nuclei; however, precursor phenomena may 
be observed even in normal nuclei if they are in the proximity 
of the critical point of the phase transition.
 The first proposal for possible evidence of a precursor was 
enhancement of the M1 cross section in proton inelastic scattering 
\cite{prl_42_1034_1979,plb_89_327_1980}.
 However, this prediction was not supported by measurements of 
${}^{12}{\rm C}(p,p'){}^{12}{\rm C}(1^+,T=1)$ 
\cite{prl_44_1656_1980,prc_21_2147_1980}.
 A possible reason for the absence of the precursor may be 
that the M1 cross section involves both pionic (spin-longitudinal)
and rho-mesonic (spin-transverse)
transitions and that the contribution from the 
rho-mesonic transition might mask the pionic effect.

 A further possible source of evidence of a precursor was 
proposed by Alberico {\it et al.} \cite{plb_92_153_1980}.
 They calculated the pionic and rho-mesonic 
response functions, $R_L$ and $R_T$, in the quasielastic 
scattering (QES) region. 
 Their results showed significant enhancement in $R_L/R_T$ 
due to nuclear spin--isospin correlations.
 Great effort has been made to extract the 
spin response functions $R_L$ and $R_T$ experimentally 
in $(\vec{p},\vec{p}\,')$ scatterings 
and $(\vec{p},\vec{n})$ reactions 
at intermediate energies \cite{ppnp_56_446_2006}.
 None of the observed ratios show evidence of the theoretically 
expected enhancement.
 The fact that the rho-mesonic response $R_T$ is equally
important in determining the ratio $R_L/R_T$
 means that the pionic enhancement may be masked by the contribution 
from the rho-mesonic component.
 Recent analysis of QES data shows pionic enhancement 
in the spin-longitudinal cross section that well represents 
$R_L$, suggesting that the lack of enhancement 
in $R_L/R_T$ is due to the rho-mesonic component 
\cite{prc_69_054609_2004}.
 It should be noted that pionic enhancement 
has been also observed in the pure pionic excitation of 
${}^{16}{\rm O}(p,p'){}^{16}{\rm O}(0^-,T=1)$ scattering 
at $T_p$ = 295 MeV \cite{plb_632_485_2006}.

 Recent progress in the development of high intensity 
polarized ion sources and high efficiency 
neutron polarimeters has enabled the measurement 
of a complete set of polarization observables 
for the ${}^{12}{\rm C}(\vec{p},\vec{n}){}^{12}{\rm N}({\rm g.s.},1^+)$ 
reaction at large momentum transfers covering the 
critical momentum $q$ $\simeq$ 1.7 ${\rm fm^{-1}}$ 
of pion condensation.
 This Gamow--Teller (GT) transition is the isobaric 
analog to the M1 excitation 
of ${}^{12}{\rm C}(p,p'){}^{12}{\rm C}(1^+,T=1)$ scattering.
In addition, the $(p,n)$ reaction is free from isospin mixing effects.
 Thus, it is very interesting to study nuclear correlation 
effects in this reaction by separating the cross section 
into pionic and rho-mesonic components with polarization 
observables.
 Furthermore, distorted wave impulse approximation (DWIA) 
calculations for finite nuclei employing nuclear correlations 
with continuum random phase approximation (RPA) are available.

 In this Letter, we present the measurement of the cross section
and a complete set of polarization observables 
for the ${}^{12}{\rm C}(\vec{p},\vec{n}){}^{12}{\rm N}({\rm g.s.},1^+)$ 
reaction.
 An incident beam energy of 296 MeV is used. This is one of the best 
energies to study GT transitions since the spin excitations 
including GT transitions are dominant in the 
$(p,n)$ reaction near 300 MeV \cite{prc_31_488_1985}.
 Furthermore, distortion effects become minimum around this 
incident energy.
 This allows us to extract nuclear structure information reliably
by the 
$(p,n)$ reaction, such as the nuclear correlation effects.
 We compare our results with DWIA calculations.
 Possible evidence of nuclear correlations is observed 
in the comparison between the experimental and theoretical results.
 We also compare our data with DWIA calculations employing 
RPA response functions including the $\Delta$ isobar 
in order to assess the nuclear correlation effects quantitatively. 

 The data were obtained with a neutron time-of-flight (NTOF) 
system \cite{nim_a369_120_1996} 
with a neutron detector and polarimeter NPOL3 system 
\cite{nim_a547_569_2005} at the Research Center for Nuclear Physics, 
Osaka University.
 The NTOF system consists of a beam-swinger dipole magnet, 
a neutron spin-rotation (NSR) magnet, and a 100-m tunnel. 
 The beam polarization was continuously monitored using two 
$\vec{p}+p$ scattering polarimeters; its typical magnitude 
was about 0.70.
 The beam energy was determined to be $296\pm 1$ MeV 
from the kinematic energy shift between two peaks 
from 
${}^{7}{\rm Li}(p,n){}^{7}{\rm Be}({\rm g.s.}+0.43\,{\rm MeV})$ and 
${}^{12}{\rm C}(p,n){}^{12}{\rm N}({\rm g.s.})$.
 In the beam-swinger system, a beam with a typical 
current of 500 nA was incident on a self-supporting 
${}^{\rm nat}{\rm C}$ (98.9\% ${}^{12}{\rm C}$) target 
with a thickness of 89 ${\rm mg/cm^2}$.
 Neutrons from the target passed through the NSR magnet 
and were measured by the NPOL3 system in the 100-m TOF tunnel 
with a resolution of about 500 keV FWHM.
 The neutron detection efficiency of NPOL3 was determined 
to be $0.025\pm0.002$ using
${}^{7}{\rm Li}(p,n){}^{7}{\rm Be}({\rm g.s.}+0.43\,{\rm MeV})$ at $0^{\circ}$, 
whose cross section is known for $T_p$ = 80--795 MeV \cite{prc_41_2548_1990}. 
 The neutron polarimetry of NPOL3 was calibrated using 
${}^{12}{\rm C}(\vec{p},\vec{n}){}^{12}{\rm N}({\rm g.s.})$ at $0^{\circ}$
\cite{nim_a547_569_2005}. 
The effective analyzing power $A_{y;{\rm eff}}$ of NPOL3
was determined to be $A_{y;{\rm eff}}=0.151\pm 0.007\pm 0.004$, 
where the first and second uncertainties are statistical and systematic 
uncertainties, respectively.

\begin{figure}
\begin{center}
\includegraphics[width=0.9\linewidth,clip]{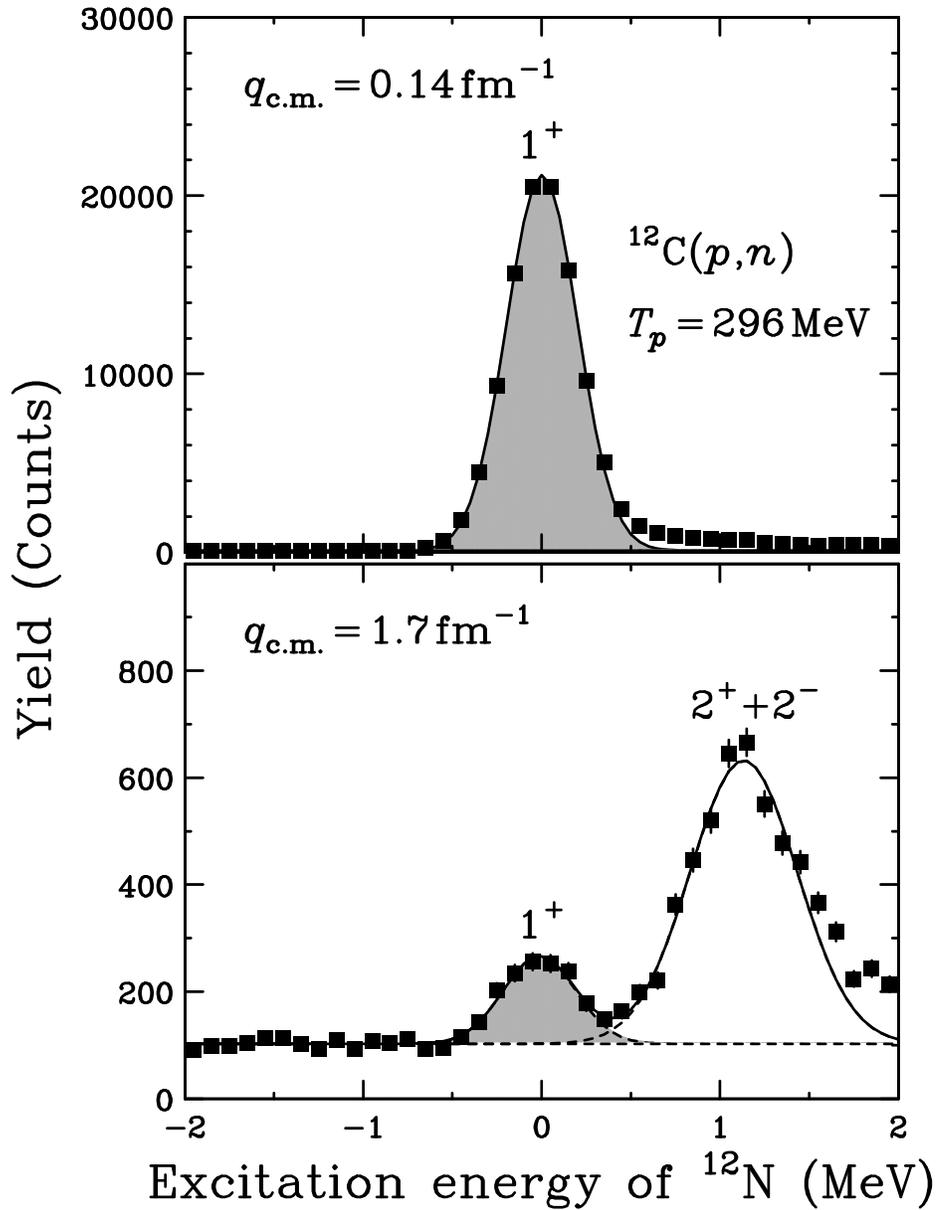}
\end{center}
\caption{
 Excitation energy spectra for 
${}^{12}{\rm C}(p,n){}^{12}{\rm N}$ at 
$T_p$ = 296 MeV and $q$ = $0.14\,\mathrm{fm^{-1}}$ (upper panel) 
and $q$ = $1.7\,\mathrm{fm^{-1}}$ (lower panel). 
 The dashed curves and straight dashed line represent 
fits to the individual peaks and background, respectively.
 The solid curve shows the sum of the peak fitting. 
\label{fig:fit}}
\end{figure}

 Figure~\ref{fig:fit} shows the excitation energy 
spectra of ${}^{12}{\rm C}(p,n){}^{12}{\rm N}$
for momentum transfers $q$ = $0.14\,\mathrm{fm}^{-1}$ 
and $1.7\,\mathrm{fm}^{-1}$.
 The GT $1^+$ state at $E_x$ = 0 MeV (ground state) forms a 
pronounced peak for $q$ = $0.14\,\mathrm{fm}^{-1}$, though it is 
not fully resolved from the neighboring states for 
$q$ = $1.7\,\mathrm{fm}^{-1}$.
 Therefore, we performed peak fitting for $E_x$ $\le$ 1.5 MeV 
to extract the yield of the $1^+$ state.
 The first and second excited states with 
$J^{\pi}$ = $2^+$ and $2^-$ at $E_x$ = 0.96 and 1.19 MeV 
\cite{npa_433_100_1985}
were considered in the peak fitting and were assumed to 
form a single peak because the present energy resolution 
could not resolve these two peaks.
 The continuum background from wrap-around 
and ${}^{13}{\rm C}(p,n)$ events is also considered in the peak fitting.
 The dashed curves in Fig.~\ref{fig:fit} represent the fits 
to the individual peaks, while the straight dashed line and 
solid curve represent the background and the sum of the peak 
fitting, respectively.
 The peak fittings at all momentum transfers 
were satisfactory for extracting the $1^+$ yield.

\begin{figure}
\begin{center}
\includegraphics[width=0.9\linewidth,clip]{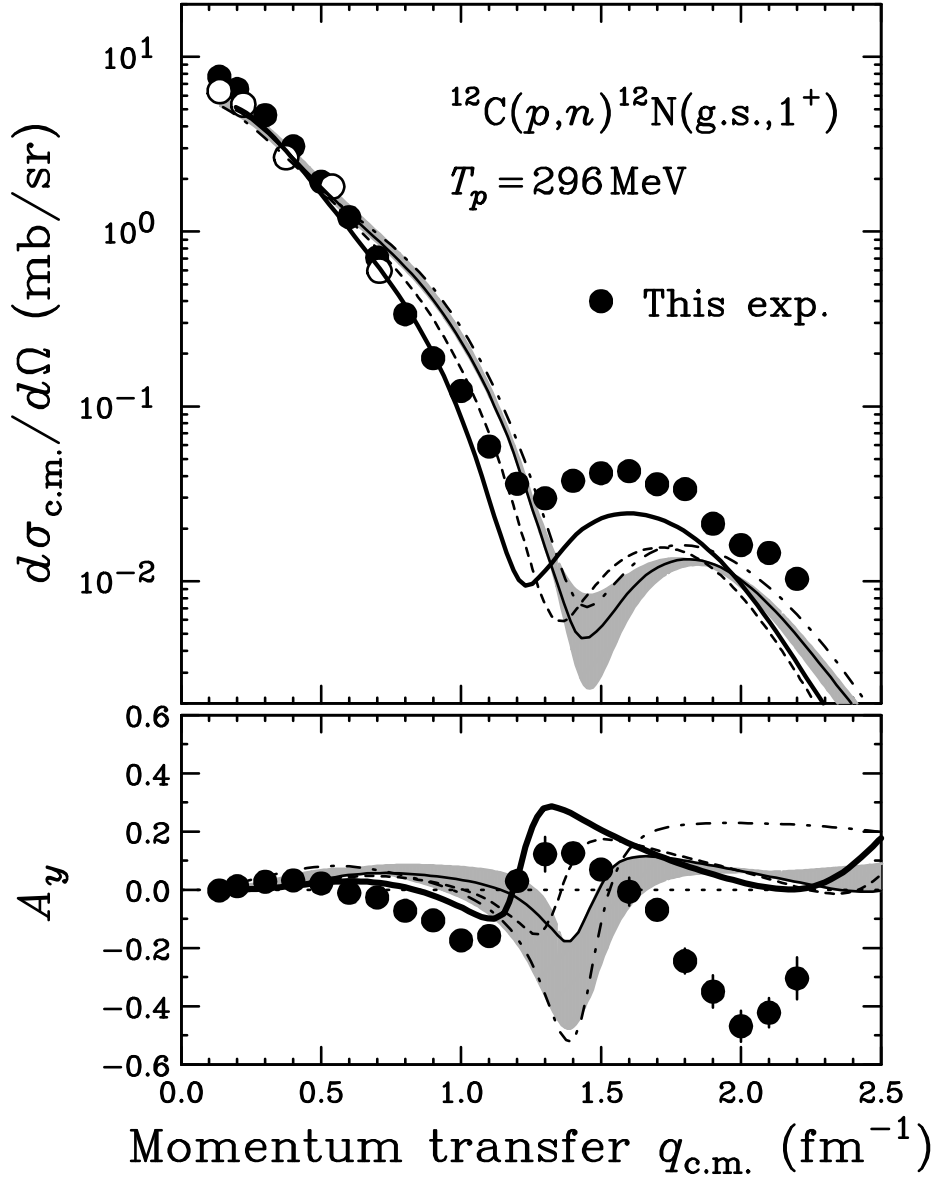}
\end{center}
\caption{
 Comparison between experimental and 
theoretical cross sections (upper panel) and 
analyzing powers (lower panel) for 
${}^{12}{\rm C}(p,n){}^{12}{\rm N}({\rm g.s.},1^+)$ at 
$T_p$ = 296 MeV.
 The thin solid curves are the DWIA calculations for a 
$0p_{1/2}0p_{3/2}^{-1}$ configuration and the bands 
represent the OMP dependence.
 The dotdash curves denote the DWIA results with CKWFs.
 The dashed and thick solid curves are, respectively, the DWIA results 
with the free response function employing $m^*(0)$ = $0.7m_N$ 
and 
with the RPA response function employing $g'_{NN}$ = 0.65, 
$g'_{N\Delta}$ = 0.35, and $m^*(0)$ = $0.7m_N$.
\label{fig:cs}}
\end{figure}

 The differential cross section for 
${}^{12}{\rm C}(p,n){}^{12}{\rm N}({\rm g.s.},1^+)$ at $T_p$ = 296 MeV
is shown in the upper panel of Fig.~\ref{fig:cs}.
 The data for the analyzing power were also measured and the 
results are displayed in the lower panel.
 The momentum-transfer dependence was measured in the range 
$q$ = 0.1--2.2 $\mathrm{fm^{-1}}$, 
covering the maximum at 
$q$ $\simeq$ $1.6\,\mathrm{fm^{-1}}$.
 The data at $T_p$ = 295 MeV \cite{prc_51_2871_1995} 
are also displayed as open circles.
 Both data sets are consistent with each other taking into account 
the statistical and systematic uncertainties.

 We performed DWIA calculations using 
the computer code {\sc crdw} \cite{prc_63_044609_2001}.
 The optical model potential (OMP) was deduced 
from the global OMPs optimized for ${}^{12}{\rm C}$ 
in the proton energy range $T_p$ = 29--1040 MeV 
\cite{prc_41_2737_1990,prc_47_297_1993}.
 The free nucleon--nucleon ({\it NN}\,) {\it t}-matrix parameters 
were taken from Bugg and Wilkin \cite{plb_152_37_1985}.
 The single particle wave functions were generated 
by a Woods--Saxon (WS) potential with 
$r_0$ = 1.27 $\mathrm{fm^{-1}}$,
$a_0$ = 0.67 $\mathrm{fm^{-1}}$ \cite{bohr_mottelson}, and 
$V_{\rm so}$ = 6.5 MeV \cite{prc_51_269_1995}.
 The depth of the WS potential was adjusted to 
reproduce the separation energies of the $0p_{3/2}$ orbits.
 The transition form factor was normalized to reproduce 
the beta-decay $ft$ value of 13178 s \cite{prc_17_280_1978}, 
whose corresponding beta-decay strength $B({\rm GT})$ was 
deduced to be $B({\rm GT})$ = 0.873 \cite{plb_349_427_1995}.
 The thin solid curves in Fig.~\ref{fig:cs} show the DWIA results.
 The normalization factor for the transition form factor 
is $N$ = 0.17. 
 These calculations reproduce the experimental data 
reasonably well for small momentum transfers of 
$q \lesssim 0.5\,{\rm fm^{-1}}$.
 However, the angular distributions of both the cross section 
and the analyzing power shift to higher momentum transfer.
 Furthermore, the calculations significantly underestimate 
the cross section at large momentum transfers 
of $q$ $\simeq$ $1.6\,\mathrm{fm^{-1}}$.

 We also investigated the OMP dependence of DWIA calculations 
by using other OMPs \cite{prc_43_2773_1991,prc_48_1106_1993}.  
The results are shown by the bands in Fig.~\ref{fig:cs}.
 The OMP dependence is clearly seen around the cross section 
minimum at $q$ $\simeq$ 1.4 $\mathrm{fm^{-1}}$.
 However, neither the discrepancy of the angular distributions 
nor the underestimation of the cross section at large momentum 
transfer can be resolved.

 We also evaluated the proton--particle and neutron--hole 
configuration dependence on the transition form factor.
 In the above calculations, the Hartree--Fock (HF) state 
of ${}^{12}{\rm C}$ was described as the state 
fully occupying the $0s_{1/2}$ and $0p_{3/2}$ orbits.
 Thus, the GT transition is described as a 
combination of a $0p_{1/2}$ proton--particle and a 
$0p_{3/2}^{-1}$ neutron--hole ($0p_{1/2}0p_{3/2}^{-1}$).
 However, the Cohen and Kurath wave functions (CKWFs) 
\cite{np_73_1_1965}
show that the 
$0p_{3/2}0p_{3/2}^{-1}$, 
$0p_{3/2}0p_{1/2}^{-1}$, and 
$0p_{1/2}0p_{1/2}^{-1}$ configurations 
also play important roles in the GT transition.
 Therefore we performed DWIA calculations with CKWFs using
the computer code {\sc dw81} \cite{dw81}.
 We used the same OMPs and single particle wave functions as 
in the previous calculations.
 The {\it NN} {\it t}-matrix parametrized 
by Franey and Love \cite{prc_31_488_1985} at 325 MeV was used.
 The dotdash curves in Fig.~\ref{fig:cs} denote the DWIA
results using the CKWSs with $N$ = 0.89.
 The cross section is slightly enhanced at all momentum transfers. 
 However, the oscillatory structures, especially 
the momentum transfers where the cross section
takes a minimum or maximum value, were not 
improved by using the realistic CKWFs.

 Therefore, we investigated the non-locality 
of the nuclear mean field by using a local effective mass
approximation \cite{prc_59_3177_1999} in the form of 
\begin{equation}
m^*(r) = m_N-\frac{f_{\rm WS}(r)}{f_{\rm WS}(0)}(m_N-m^*(0)),
\label{eq:effmass}
\end{equation}
where $m_N$ is the nucleon mass and $f_{\rm WS}(r)$ is a 
WS radial form.
 The dashed curves in Fig.~\ref{fig:cs} show the DWIA 
results with $m^*(0)$ = $0.7m_N$ 
\cite{plb_126_421_1983,npa_481_381_1988} and $N$ = 0.17.
 The angular distributions of both the cross section 
and the analyzing power shift to lower momentum transfer 
for smaller $m^*(0)$, 
and a standard value of $m^*(0)\simeq 0.7m_N$ 
improves the agreement with the data.
 This is because the transition form factor 
effectively moves outwards due to the Perey factor effect
\cite{npa_32_353_1962}.
 However, there is still a discrepancy between the experimental 
and theoretical results around $q$ $\simeq$ 
$1.6\,\mathrm{fm^{-1}}$.

 Next, we consider the nuclear correlation effects.
 We performed DWIA calculations with the RPA
response functions employing
the $\pi+\rho+g'$ model interaction $V_{\rm eff}$.
 In the spin--isospin channel, $V_{\rm eff}$ can be 
expressed as \cite{ppnp_56_446_2006}
\begin{equation}
V_{\rm eff}(\bm{q},\omega)
= V_{L}(\bm{q},\omega)+V_{T}(\bm{q},\omega)\ , \label{eq:Veff}
\end{equation}
with spin-longitudinal and spin-transverse parts
\begin{equation}
\begin{split}
V_{L}(\bm{q},\omega) 
&= W_{L}^{NN}(q,\omega)\, 
       (\bm{\tau}_1\cdot\bm{\tau}_2)
       (\bm{\sigma}_1\cdot{\hat{\bm{q}}})
       (\bm{\sigma}_2\cdot{\hat{\bm{q}}})   \\
&+ W_{L}^{N\Delta}(q,\omega)
        \left[
        \left\{(\bm{\tau}_1\cdot\bm{T}_2)
               (\bm{\sigma}_1\cdot{\hat{\bm{q}}})
               (\bm{S}_2     \cdot{\hat{\bm{q}}})
        + (1\leftrightarrow 2)\right\} + {\rm h.c.} 
        \right]                             \\
&+ W_{L}^{\Delta\Delta}(q,\omega)\, 
        \left[
    \left\{
     (\bm{T}_1\cdot\bm{T}_2^{\dagger})
     (\bm{S}_1          \cdot{\hat{\bm{q}}})
     (\bm{S}_2^{\dagger}\cdot{\hat{\bm{q}}}) + 
     (\bm{T}_1\cdot\bm{T}_2)
     (\bm{S}_1\cdot{\hat{\bm{q}}})
     (\bm{S}_2\cdot{\hat{\bm{q}}})
    \right\} + {\rm h.c.} 
    \right]\ ,                              \\
V_{T}(\bm{q},\omega) 
&= W_{T}^{NN}(q,\omega)\, 
    (\bm{\tau}_1\cdot\bm{\tau}_2)
    (\bm{\sigma}_1\times{\hat{\bm{q}}})
    (\bm{\sigma}_2\times{\hat{\bm{q}}})     \\
&+ W_{T}^{N\Delta}(q,\omega)
    \left[
    \left\{
    (\bm{\tau}_1\cdot\bm{T}_2)
    (\bm{\sigma_1}\times{\hat{\bm{q}}})
    (\bm{S}_2     \times{\hat{\bm{q}}})
    + (1\leftrightarrow 2) 
    \right\} + {\rm h.c.} 
    \right]                                 \\
&+ W_{T}^{\Delta\Delta}(q,\omega)\, 
    \left[
    \left\{
     (\bm{T}_1\cdot\bm{T}_2^{\dagger})
     (\bm{S}_1          \times{\hat{\bm{q}}})
     (\bm{S}_2^{\dagger}\times{\hat{\bm{q}}}) + 
     (\bm{T}_1\cdot\bm{T}_2)
     (\bm{S}_1\times{\hat{\bm{q}}})
     (\bm{S}_2\times{\hat{\bm{q}}})
    \right\} + {\rm h.c.} 
    \right]\ ,        
\end{split}
\label{eq:veffective}
\end{equation}
where $W_L$ and $W_T$ are the spin-longitudinal and spin-transverse
strengths, respectively,
$\bm{\sigma}$ and $\bm{\tau}$ are the nucleon Pauli spin and isospin 
matrixes, and 
$\bm{S}$ and $\bm{T}$ are the spin  and isospin transition operators 
that excite {\it N} to $\Delta$.
 The strengths $W_L$ and $W_T$ are determined by the 
pion and rho-meson exchange interactions and 
the Landau--Migdal (LM) interaction $V_{\rm LM}$ specified by the LM
parameters $g'_{NN}$, $g'_{N\Delta}$, and $g'_{\Delta\Delta}$ 
\cite{ppnp_56_446_2006}.
 For the pion and rho-meson exchange interactions, we have used the coupling constants and meson parameters 
from a Bonn potential which  
treats $\Delta$ explicitly \cite{pr_149_1_1987}.
 The LM parameters were estimated to be 
$g'_{NN}$ = $0.65\pm 0.15$ and $g'_{N\Delta}$ = $0.35\pm 0.15$
\cite{prc_72_067303_2005} by using the peak position of the GT 
giant resonance and the GT quenching factor at $q$ = 0 
\cite{prc_55_2909_1997,plb_615_193_2005} as well as 
the isovector spin-longitudinal polarized cross section in 
the QES process at $q$ $\simeq$ 1.7 ${\rm fm^{-1}}$ 
\cite{prc_69_054609_2004}.
 The thick solid curves in Fig.~\ref{fig:cs} show the 
DWIA result using $g'_{NN}$ = 0.65, $g'_{N\Delta}$ = 0.35, 
and $m^*(0)$ = $0.7m_N$ with $N$ = 0.28.
 Here, we fixed $g'_{\Delta\Delta}$ = 0.5 \cite{prc_23_1154_1981}
since the $g'_{\Delta\Delta}$ dependence of the results is very weak.
 In the continuum RPA, the GT state couples to 
particle-unbound $1^+$ states, which shifts the response 
function in the coordinate space ($r$-space) to large $r$ values.
 Thus the angular distributions further shift to 
lower momentum transfer.
 The RPA correlation also enhances the cross section 
at large momentum transfers of $q$ $\simeq$ 1.6 ${\rm fm^{-1}}$.
 Consequently, the discrepancy between the experimental 
and theoretical 
results could be resolved in part by considering the nuclear 
correlation effects in the RPA together with the non-locality effects
of the nuclear mean field.
 Nuclear correlation effects are expected to be more clearly 
seen in the oscillatory structures of polarized cross sections.
 Therefore, we have separated the cross section into 
polarized cross sections by using the measured polarization 
observables.

\begin{figure}
\begin{center}
\includegraphics[width=0.9\linewidth,clip]{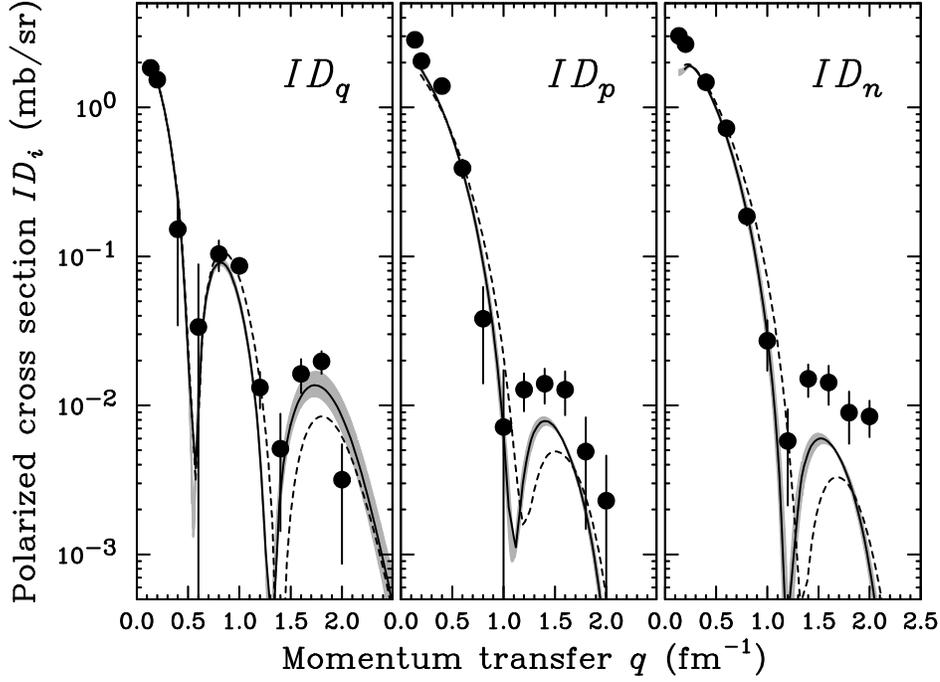}
\end{center}
\caption{
 Comparison between experimental and theoretical 
$ID_q$ (left panel), $ID_p$ (middle panel), and 
$ID_n$ (right panel)  for 
${}^{12}{\rm C}(p,n){}^{12}{\rm N}({\rm g.s.},1^+)$ at 
$T_p$ = 296 MeV.
 The solid curves are the DWIA calculations with the RPA response 
functions using $(g'_{NN},g'_{N\Delta},g'_{\Delta\Delta})$ = 
$(0.65,0.35,0.50)$ and $m^*(0)$ = $0.7m_N$.
 The bands represent the $g'_{NN}$ and $g'_{N\Delta}$ 
dependences of the DWIA results within 
$0.50\le g'_{NN}\le 0.80$ and 
$0.20\le g'_{N\Delta}\le 0.50$.
 The dashed curves show the DWIA results with the free response 
functions using $m^*(0)$ = $0.7m_N$.
\label{fig:idi}}
\end{figure}

\begin{figure}
\begin{center}
\includegraphics[width=0.9\linewidth,clip]{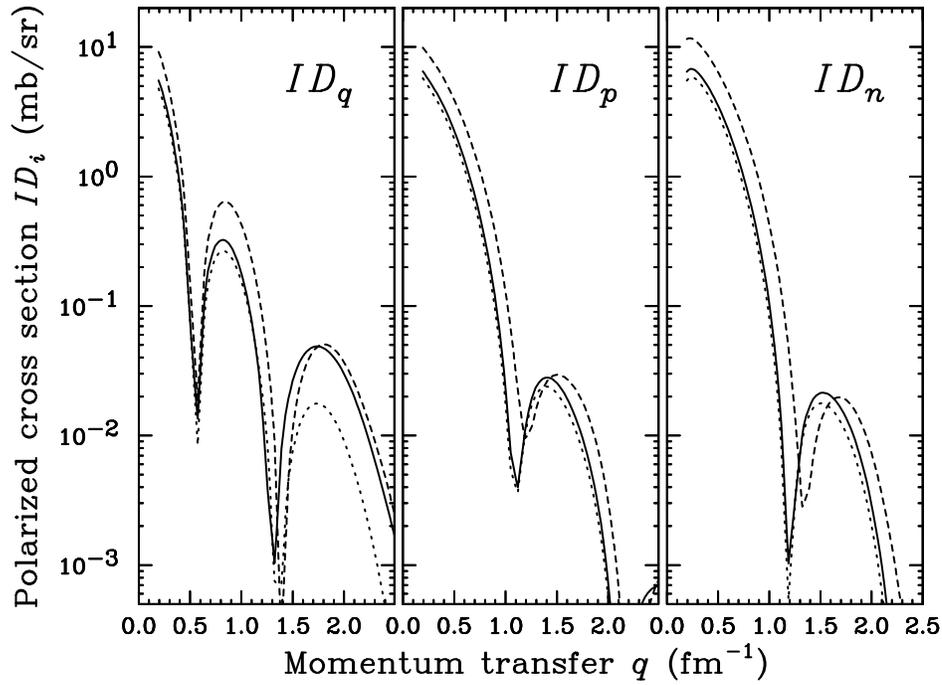}
\end{center}
\caption{
 Theoretical calculations without normalization for 
$ID_q$ (left panel), $ID_p$ (middle panel), and 
$ID_n$ (right panel)  for 
${}^{12}{\rm C}(p,n){}^{12}{\rm N}({\rm g.s.},1^+)$ at 
$T_p$ = 296 MeV.
 The dashed curves show the DWIA results with the free response 
function.
 The dotted and solid curves are the DWIA calculations with 
the RPA response functions employing the Landau--Migdal 
interaction ($g'$) only and the $\pi+\rho+g'$ model 
interaction, respectively.
\label{fig:rpa}}
\end{figure}

 Figure~\ref{fig:idi} shows three polarized cross sections, 
$ID_q$, $ID_p$, and $ID_n$, as a function of momentum transfer.
 The spin-scalar polarized cross section $ID_0$ 
is not shown because it is very small in the present 
spin-flip GT case.
 The error bars represent the statistical uncertainties 
of the data.
 The systematic uncertainties are  
about 30\% for all $ID_i$ at $q$ $\simeq$ 1.6 $\mathrm{fm^{-1}}$.
 The solid curves in Fig.~\ref{fig:idi} denote the DWIA calculations 
with the RPA response function, and the bands 
represent the $g'_{NN}$ and $g'_{N\Delta}$ dependences 
within 
$g'_{NN}$ = $0.65\pm 0.15$ and $g'_{N\Delta}$ = $0.35\pm 0.15$.
 The dashed curves are the DWIA calculations with the 
free response function.
 The momentum transfer dependence of all the three 
spin-dependent $ID_i$ shift 
to lower momentum transfer and are enhanced around 
$q$ = 1.6 ${\rm fm^{-1}}$ due to 
the nuclear correlation effects,  
which improves the agreement with the data.
 Figure~\ref{fig:rpa} shows the DWIA calculations 
without the normalization.
 The dashed and dotted curves are the DWIA results with the 
free response function and with the RPA response function 
employing only the LM interaction, respectively.
 The figure shows quenching of all $ID_i$ over the whole momentum 
transfer region due to the repulsive LM interaction,
as well as forward shifts in the angular distributions
due to outward shifts of the response functions in 
$r$-space by the continuum coupling in the RPA.
 The solid curves are the DWIA results with the 
RPA response function employing the $\pi+\rho+g'$ model 
interaction $V_{\rm eff}$.
 The pion exchange effects in $ID_q$ are clearly seen as 
an enhancement at large momentum transfers.
 In contrast, the rho-meson exchange 
effects in $ID_p$ and $ID_n$ are expected to be small, 
mainly due to the weak momentum transfer dependence 
of $W_T$ in the present momentum transfer region 
\cite{ppnp_56_446_2006}.
 It should be noted that 
in the analysis shown in Fig.~\ref{fig:idi}, 
the transition form factors are normalized to reproduce 
$B({\rm GT})$, and thus the calculated $ID_i$ are normalized
at small momentum transfers.
 Therefore, the quenching due to the LM interaction is 
effectively included through the normalization factor 
$N$ at small momentum transfers.
 However, the modification of the angular distributions
due to the shape change of the response functions in $r$-space 
could not be included through this 
normalization, and thus this effect is observed as an 
enhancement in $ID_p$ and $ID_n$ in Fig.~\ref{fig:idi} at 
large momentum transfers of $q$ $\simeq$ 1.5 $\mathrm{fm^{-1}}$.
 We also note that the momentum transfer dependence 
of the {\it NN} {\it t}-matrix is important for the 
momentum transfer dependence of $ID_i$.
 For example, the first minimum of $ID_q$ is due to the 
first minimum of the relevant {\it NN} {\it t}-matrix 
component \cite{prc_31_488_1985}.

\begin{figure}
\begin{center}
\includegraphics[width=0.9\linewidth,clip]{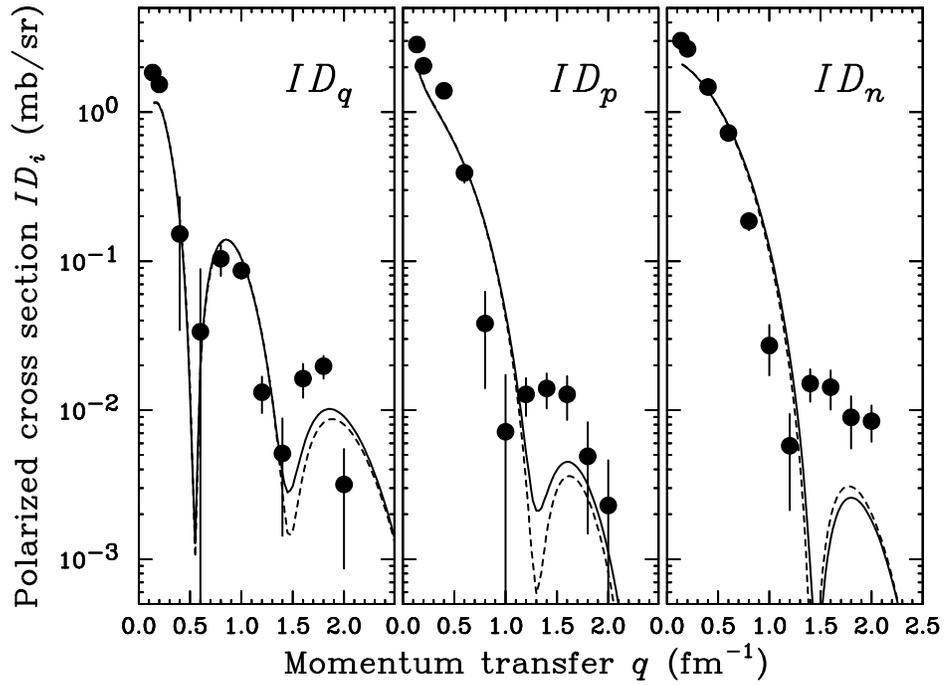}
\end{center}
\caption{
 DWIA predictions for 
${}^{12}{\rm C}(p,n){}^{12}{\rm N}({\rm g.s.},1^+)$ at 
$T_p$ = 296 MeV.
 The solid curves are the DWIA results with CKWFs.
 The dashed curves represent the DWIA results 
for a $0p_{1/2}0p_{3/2}^{-1}$ configuration.
 The experimental data are same as those in Fig.~\protect{\ref{fig:idi}}.
\label{fig:idick}}
\end{figure}

 The DWIA calculations employing nuclear correlation 
effects give better descriptions of all three spin-dependent 
$ID_i$. However, there are still significant discrepancies
at large momentum transfers around 
$q$ $\simeq$ 1.6 ${\rm fm^{-1}}$.
 The under-estimation of the theoretical calculations might be 
resolved using a realistic HF state of ${}^{12}{\rm C}$
because the DWIA calculation with CKWFs is enhanced 
around $q$ $\simeq$ 1.6 ${\rm fm^{-1}}$ compared with that for the 
pure $0p_{1/2}0p_{3/2}^{-1}$ configuration, as shown in 
Fig.~\ref{fig:cs}.
 Thus, we deduced $ID_i$ from the DWIA results with 
CKWFs; the results are shown by the solid curves in Fig.~\ref{fig:idick}.
 The dashed curves correspond to the DWIA results 
for the pure $0p_{1/2}0p_{3/2}^{-1}$ configuration.
 The use of realistic CKWFs enhances $ID_q$ and $ID_p$ 
at large momentum transfers, 
and thus the discrepancies between the experimental and theoretical 
results seen in Fig.~\ref{fig:idi} would be 
due to the simplification of the HF
state of ${}^{12}{\rm C}$ in the RPA calculations.
 However, the calculations with CKWFs give 
smaller $ID_n$ values.
 Thus, the reason for the difference in $ID_n$ seems to be 
different to that for $ID_q$ and $ID_p$.
 This is possibly due to medium modifications 
of the effective {\it NN}\, interaction.
 It should be noted that the enhancement of $ID_n$ 
at large momentum transfers
is commonly seen in QES and in stretched state excitations 
\cite{plb_645_402_2007} 
and that this enhancement indicates the enhancement 
of $B$ of the {\it NN} scattering amplitude 
in a Kerman--MacManus--Thaler (KMT) representation 
\cite{ap_8_551_1959}.
 Thus, the enhancement of $B$ might be 
responsible for the enhancement of the 
corresponding rho-mesonic $ID_n$.
 We note that in both QES and stretched state excitations,
no modification of the spin-transverse amplitude $F$ 
has been observed for $ID_p$, whereas 
a small reduction of the spin-longitudinal amplitude $E$ 
has been indicated for $ID_q$ \cite{plb_645_402_2007}.
 If we take into account the reduction of $E$,  
the enhancement of $ID_q$ due to the 
nuclear correlations
should be larger than the current prediction using 
$g'_{NN}$ = 0.65 and $g'_{N\Delta}$ = 0.35, which 
can be achieved 
by using smaller, reasonable LM parameters, 
$g'_{NN}\simeq 0.6$ and $g'_{N\Delta}\simeq 0.2$ 
\cite{ppnp_56_446_2006}.

 In conclusion, our measurement of 
the cross section and a complete set of 
polarization observables of 
the 
${}^{12}{\rm C}(\vec{p},\vec{n}){}^{12}{\rm N}(g.s.,1^+)$
reaction enables us to study nuclear correlation 
effects at an intermediate energy of 
$T_p$ = 296 MeV where the theoretical DWIA calculations 
should be reliable due to the simple reaction mechanism.
 A significant difference in the momentum-transfer 
dependence and an enhancement around $q$ $\simeq$ 1.6 
$\mathrm{fm^{-1}}$
were observed in the cross section
compared to the DWIA calculations 
with Cohen--Kurath wave functions.
 A DWIA calculation employing the RPA response function 
with $g'_{NN}$ = 0.65 and $g'_{N\Delta}$ = 0.35 and
a local effective mass with 
$m^*(0)$ = $0.7m_N$ reproduces the momentum-transfer 
dependence and gives an enhancement of the cross section 
at $q$ $\simeq$ 1.6 $\mathrm{fm^{-1}}$.
 All three spin-dependent polarized cross sections also
support the existence of nuclear correlation effects.
 In theoretical calculations, the use of a realistic 
shell-model wave function
seems to be important to reproduce the experimental data
at large momentum transfers. However, this effect 
could not be included in the present RPA calculations.
 Therefore, a more comprehensive and detailed 
theoretical analysis is needed to explain quantitatively
the nuclear correlations inside nuclei.

\ack
 We thank the RCNP cyclotron crew for providing a good 
quality beam for our experiments.
 This work was supported in part by the Grants-in-Aid for
Scientific Research Nos.~14702005 and 16654064
of the Ministry of Education, Culture, Sports, 
Science, and Technology of Japan.

\clearpage

\clearpage

\clearpage

\clearpage

\clearpage

\clearpage

\bibliographystyle{elsart-num}
\bibliography{wakasa}

\begin{thebibliography}{10}
\expandafter\ifx\csname url\endcsname\relax
  \def\url#1{\texttt{#1}}\fi
\expandafter\ifx\csname urlprefix\endcsname\relax\def\urlprefix{URL }\fi

\bibitem{Migdal1}
A. B. Migdal, Zh. Eksp. Teor. Fiz. 61, (1971) 2210; Sov. Phys. JETP 34, (1972)
  1184.

\bibitem{Chandra}
P.~O. Slane, D.~J. Helfand, S.~S. Murray, Astrophys.\ J. 571 (2002) L45.

\bibitem{tsuruta}
S.~Tsuruta, et~al., Astrophys.\ J. 571 (2002) L143.

\bibitem{prl_42_1034_1979}
H.~Toki, W.~Weise, Phys.\ Rev.\ Lett. 42 (1979) 1034.

\bibitem{plb_89_327_1980}
J.~Delorme, et~al., Phys.\ Lett.\ B 89 (1980) 327.

\bibitem{prl_44_1656_1980}
J.~R. Comfort, W.~G. Love, Phys.\ Rev.\ Lett. 44 (1980) 1656.

\bibitem{prc_21_2147_1980}
J.~R. Comfort, et~al., Phys.\ Rev.\ C 21 (1980) 2147.

\bibitem{plb_92_153_1980}
W.~M. Alberico, M.~Ericson, A.~Molinari, Phys.\ Lett.\ B 92 (1980) 153.

\bibitem{ppnp_56_446_2006}
M.~Ichimura, H.~Sakai, T.~Wakasa, Prog.\ Part.\ Nucl.\ Phys. 56 (2006) 446.

\bibitem{prc_69_054609_2004}
T.~Wakasa, et~al., Phys.\ Rev.\ C 69 (2004) 054609.

\bibitem{plb_632_485_2006}
T.~Wakasa, et~al., Phys.\ Lett.\ B 632 (2006) 485.

\bibitem{prc_31_488_1985}
M.~A. Franey, W.~G. Love, Phys.\ Rev.\ C 31 (1985) 488.

\bibitem{nim_a369_120_1996}
H.~Sakai, et~al., Nucl.\ Instrum.\ Methods Phys.\ Res.\ A 369 (1996) 120.

\bibitem{nim_a547_569_2005}
T.~Wakasa, et~al., Nucl.\ Instrum.\ Methods Phys.\ Res.\ A 547 (2005) 569.

\bibitem{prc_41_2548_1990}
T.~N. Taddeucci, et~al., Phys.\ Rev.\ C 41 (1990) 2548.

\bibitem{npa_433_100_1985}
F.~Ajzenberg-Selove, et~al., Nucl.\ Phys.\ A 433 (1985) 100.

\bibitem{prc_51_2871_1995}
T.~Wakasa, et~al., Phys.\ Rev.\ C 51 (1995) R2871.

\bibitem{prc_63_044609_2001}
K.~Kawahigashi, et~al., Phys.\ Rev.\ C 63 (2001) 044609.

\bibitem{prc_41_2737_1990}
S.~Hama, et~al., Phys.\ Rev.\ C 41 (1990) 2737.

\bibitem{prc_47_297_1993}
E.~D. Cooper, et~al., Phys.\ Rev.\ C 47 (1993) 297.

\bibitem{plb_152_37_1985}
D.~V. Bugg, C.~Wilkin, Phys.\ Lett.\ B 152 (1985) 37.

\bibitem{bohr_mottelson}
A.~Bohr, B.~R. Mottelson, Nuclear structure Volume I: Single-Particle Motion,
  Benjamin, New York, 1969.

\bibitem{prc_51_269_1995}
K.~Nishida, M.~Ichimura, Phys.\ Rev.\ C 51 (1995) 269.

\bibitem{prc_17_280_1978}
D.~E. Alburger, A.~M. Nathan, Phys.\ Rev.\ C 17 (1978) 280.

\bibitem{plb_349_427_1995}
K.~Schreckenbach, et~al., Phys.\ Lett.\ B 349 (1995) 427.

\bibitem{prc_43_2773_1991}
S.~Qing-biao, F.~Da-chun, Z.~Yi-zhong, Phys.\ Rev.\ C 43 (1991) 2773.

\bibitem{prc_48_1106_1993}
F.~T. Baker, et~al., Phys.\ Rev.\ C 48 (1993) 1106.

\bibitem{np_73_1_1965}
S.~Cohen, D.~Kurath, Nucl.\ Phys. 73 (1965) 1.

\bibitem{dw81}
R. Schaeffer and J. Raynal, Program {\sc dw70} (unpublished); J. Raynal, Nucl.
  Phys. A 97, 572 (1967); J. R. Comfort, Extended version {\sc dw81}
  (unpublished).

\bibitem{prc_59_3177_1999}
T.~Wakasa, et~al., Phys.\ Rev.\ C 59 (1999) 3177.

\bibitem{plb_126_421_1983}
N.~V. Giai, P.~V. Thieu, Phys.\ Lett.\ B 126 (1983) 421.

\bibitem{npa_481_381_1988}
C.~Mahaux, R.~Sartor, Nucl.\ Phys.\ A 481 (1988) 381.

\bibitem{npa_32_353_1962}
F.~G. Perey, B.~Buck, Nucl.\ Phys.\ A 32 (1962) 353.

\bibitem{pr_149_1_1987}
R.~Machleidt, K.~Holinde, C.~Elster, Phys.\ Rep. 149 (1987) 1.

\bibitem{prc_72_067303_2005}
T.~Wakasa, M.~Ichimura, H.~Sakai, Phys.\ Rev.\ C 72 (2005) 067303.

\bibitem{prc_55_2909_1997}
T.~Wakasa, et~al., Phys.\ Rev.\ C 55 (1997) 2909.

\bibitem{plb_615_193_2005}
K.~Yako, et~al., Phys. Lett. B 615 (2005) 193.

\bibitem{prc_23_1154_1981}
W.~H. Dickhoff, et~al., Phys.\ Rev.\ C 23 (1981) 1154.

\bibitem{plb_645_402_2007}
T.~Wakasa, et~al., Phys.\ Lett.\ B 645 (2007) 402.

\bibitem{ap_8_551_1959}
A.~K. Kerman, H.~McManus, R.~M. Thaler, Ann.\ Phys. 8 (1969) 551.

\end{thebibliography}

\end{document}